\documentclass[10pt,conference]{IEEEtran}
\IEEEoverridecommandlockouts
\usepackage{cite}
\usepackage{amsmath,amssymb,amsfonts}
\usepackage{algorithmic}
\usepackage{graphicx}
\usepackage{textcomp}
\usepackage{xcolor}
\def\BibTeX{{\rm B\kern-.05em{\sc i\kern-.025em b}\kern-.08em
    T\kern-.1667em\lower.7ex\hbox{E}\kern-.125emX}}

\usepackage{booktabs}
\usepackage[hidelinks]{hyperref}
\usepackage{multirow}
\usepackage{mdframed}
\usepackage[normalem]{ulem}
\usepackage{tikz}
\usepackage{xspace}

\newcommand*\circled[1]{\tikz[baseline=(char.base)]{
            \node[shape=circle,draw,inner sep=1pt] (char) {#1};}}

\newcommand{\code}[1]{\texttt{#1}}

\newcommand{\edit}[1]{#1}

\newcommand{\improve}[1]{{\textcolor{red}{\scriptsize{$#1\uparrow$}}}}
\newcommand{\incoderb}{\textsc{InCoder}-6B\xspace}
\newcommand{\codegenb}{\textsc{CodeGen}-\textsc{Mono}-16B\xspace}
\newcommand{\davincitwo}{\textsc{code-davinci-002}\xspace}
\newcommand{\deepseek}{\textsc{DeepSeek-Coder-6.7B}\xspace}

\newcommand{\pass}[1]{pass@{$#1$}}

\newcommand{\ping}{\textsc{PInG}\xspace}

\newcounter{finding}
\newcommand{\finding}[1]{\refstepcounter{finding}
	\begin{mdframed}[linecolor=gray,roundcorner=12pt,backgroundcolor=gray!15,linewidth=3pt,innerleftmargin=2pt, leftmargin=0cm,rightmargin=0cm,topline=false,bottomline=false,rightline=false]
	\textbf{Finding \arabic{finding}:} #1
	\end{mdframed}
}

\begin{document}

\title{Enhancing Code Generation via Bidirectional Comment-Level Mutual Grounding}

\author{
  \IEEEauthorblockN{
    Yifeng Di,
    Tianyi Zhang
  }

  \IEEEauthorblockA{
    Purdue University, West Lafayette, IN, USA
  }

  \IEEEauthorblockA{
    di5@purdue.edu, tianyi@purdue.edu
  }
}

\maketitle

\begin{abstract}

Large Language Models (LLMs) have demonstrated unprecedented capability in code generation. However, LLM-generated code is still plagued with a wide range of functional errors, especially for complex programming tasks that LLMs have not seen before. Recent studies have shown that developers often struggle with inspecting and fixing incorrect code generated by LLMs, diminishing their productivity and trust in LLM-based code generation. Inspired by the mutual grounding theory in communication, we propose an interactive approach that leverages code comments as a medium for developers and LLMs to establish a shared understanding. Our approach facilitates iterative grounding by interleaving code generation, inline comment generation, and contextualized user feedback through editable comments to align generated code with developer intent. We evaluated our approach on two popular benchmarks and demonstrated that our approach significantly improved multiple state-of-the-art LLMs, e.g., \edit{17.1}\% pass@1 improvement for code-davinci-002 on HumanEval. Furthermore, we conducted a user study with 12 participants in comparison to two baselines: (1) interacting with GitHub Copilot, and (2) interacting with a multi-step code generation paradigm called Multi-Turn Program Synthesis. Participants completed the given programming tasks 16.7\% faster and with 10.5\% improvement in task success rate when using our approach. Both results show that interactively refining code comments enables the collaborative establishment of mutual grounding, leading to more accurate code generation and higher developer confidence.

\end{abstract}


\section{Introduction}

The quest for automated code generation, dating back to the 1960s~\cite{waldinger1969prow, summers1977methodology}, has evolved significantly. This field has transitioned from early deductive program synthesis methods~\cite{burstall1977transformation, manna1979synthesis, manna1980deductive, smith1985top} to the recent advent of Large Language Models (LLMs)~\cite{chen2021evaluating, li2022competition, li2023starcoder}. Despite the significant progress, LLMs often fail to align with developer intent due to factors such as reliance on spurious features, lack of user context, and misunderstanding of complex specifications~\cite{kou2024large, vaithilingam2022expectation, yu2024codereval, wang2024coderag, wang2024large}. Recent efforts to address these limitations include model fine-tuning~\cite{haluptzok2022language, chen2023improving}, new prompting strategies~\cite{wang2022no, choi2023codeprompt}, and iterative refinement paradigms~\cite{chen2023teaching, dong2023self}. However, the improvement brought by these methods is still limited. For example, Self-Debug~\cite{chen2023teaching} only achieved a 4.8\% \pass{1} increase for Codex on MBPP when test execution feedback is not available.

Recent studies~\cite{vaithilingam2022expectation, bird2022taking, barke2023grounded} highlight the critical role of bi-directional communication between developers and LLMs in programming tasks. 
While developers can edit prompts, LLMs often treat such edits as new prompts, hindering their ability to understand and incorporate developers' refinement intent. Conversational models like ChatGPT offer multi-turn dialogues for feedback. However, effectively encoding historical utterances and contextualizing feedback within conversations remains a complex and challenging task~\cite{liu2016not, tian2017make, sankar2019neural, zheng2019enhancing}.

\begin{figure}
    \centering
    \includegraphics[width=0.9\linewidth]{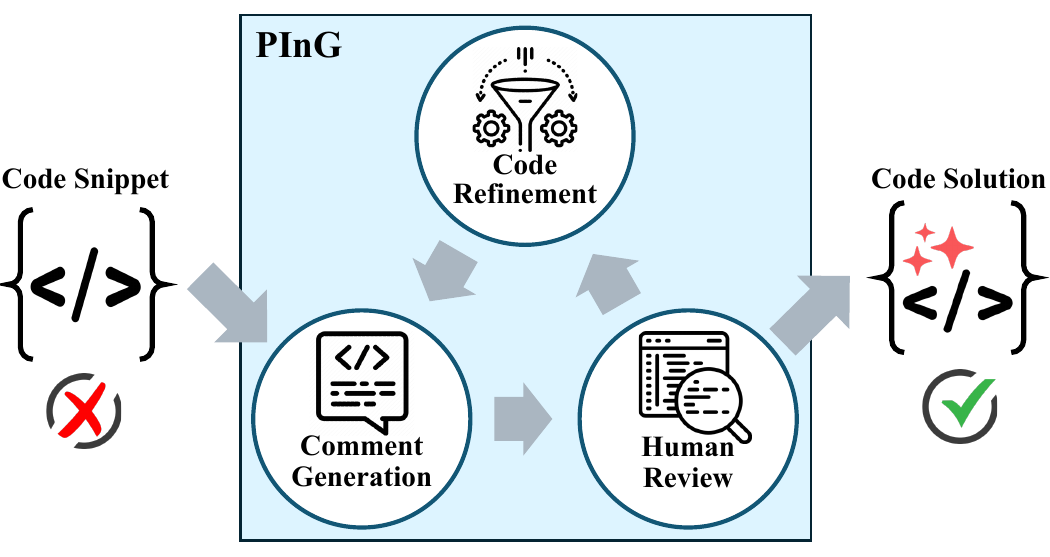}
    \caption{Generating a code solution in our pipeline.}
    \label{fig:idea}
\end{figure}

In this work, we propose a new interactive approach called Programming with Interactive Grounding (\ping). Figure \ref{fig:idea} illustrates this approach. \ping employs inline comments as a medium for bi-directional communication between a developer and the model. This approach is inspired by the grounding theory in communication~\cite{resnick1991perspectives}, which underscores the importance of mutual understanding in collaborative interactions. Unlike prior work that relies on coarse-grained code explanations for refinement~\cite{chen2023teaching, jiang2023self}, \ping's use of inline comments offers a more fine-grained approach. These comments directly address individual statements of the code, making it easier to target specific code segments for precise feedback and refinement.

Given a code snippet generated by an LLM, \ping uses a code comment generation model to create inline comments that clarify each statement's behavior. These comments provide developers with an immediate, understandable code description, helping them quickly spot potential errors. Developers can edit the comments to specify the correct behavior for the erroneous statement. A code refinement model then regenerates the statement and subsequent segment identified by the feedback, rather than the entire code snippet.

Inline comments are essentially natural language descriptions, which are suitable for developers at all levels. This approach allows for more precise error identification compared to prompt editing or conversational models, which often require the model to infer the location of errors throughout the entire code snippet. Comment editing also directly indicates error locations, leading to more efficient model-based code refinement. This simplifies and improves the process of code refinement compared to previous approaches that require regenerating the entire code snippet.

We evaluate \ping through simulated and real user studies on multiple code generation models. In our simulated user studies, we benchmark \ping against eight state-of-the-art code generation and refinement techniques~\cite{yao2022react, yao2023tree, wang2023rap, zhang2023self, jiang2023self, chen2023teaching, li2023enabling, le2023codechain}, demonstrating that \ping significantly outperforms these baselines. On HumanEval, \pass{1} rates increased by \edit{17.1}\% for code-davinci-002, \edit{9.7}\% for InCoder, and \edit{11.6}\% for CodeGen. MBPP benchmark also saw notable gains. In a user study with 12 real programmers, \ping outperformed GitHub Copilot~\cite{copilot2023} and Multi-Turn Program Synthesis~\cite{nijkamp2022codegen} in task success rate by 16.7\% and 58.3\%, respectively, while improving the task completion speed by 10.5\% and 22.9\%, respectively. Besides, participants reported 20\% increased confidence and satisfaction, demonstrating \ping's effectiveness in enhancing code quality and user experience.

In summary, our main contributions include:

\begin{enumerate}

\item We propose \ping, a new interactive code generation paradigm that enhances developer-model grounding via comment refinement.

\item We implement \ping as a VSCode extension and have open-sourced our code and data.\footnote{\label{note1}\href{https://github.com/NecoraNyaru/PInG}{https://github.com/NecoraNyaru/PInG}}

\item We comprehensively evaluate \ping with various models, benchmarks, and user studies, showing significant improvements in code accuracy and developer productivity over state-of-the-art methods.

\end{enumerate}

\section{Background} \label{sec:background}

Effective human interaction requires establishing \textit{common ground}---mutual knowledge, beliefs, and assumptions~\cite{resnick1991perspectives}. Grounding in communication involves a collaborative process~\cite{clark1989contributing}, where speakers design utterances for listener comprehension, and listeners provide feedback to demonstrate understanding~\cite{clark1982audience}. Clark and Schaefer's contribution model~\cite{clark1989contributing} outlines this as a presentation by the speaker and acceptance by the listener, supported by mechanisms including \textit{repetition, reformulation, and elaboration}~\cite{brennan1996conceptual}. The importance of common ground is underscored when communication falters due to participants' differing backgrounds~\cite{isaacs1990ostensible}.

Recent studies in code generation reveal challenges due to poor bi-directional communication between models and developers~\cite{vaithilingam2022expectation, bird2022taking, barke2023grounded}. Models often misinterpret developer intent, while developers struggle to understand the generated code~\cite{vaithilingam2022expectation}. This communication gap, stemming from the lack of common ground, leads to code that does not align with developers' expectations and hinders effective feedback~\cite{barke2023grounded}. Thus, fostering shared understanding between models and developers is a key to enhancing code generation accuracy.

In NLP, there have been some recent investigations on applying grounding theory to text-generation tasks~\cite{zhang2019dialogpt, chandu2021grounding}. Zhang et al.\cite{zhang2019dialogpt} proposed a retrieval-augmented generation model that retrieves related examples to ground the generative context in the relevant information. Chandu et al.\cite{chandu2021grounding} analyzed coordination and constraints in NLP tasks, proposing adaptations for grounding. However, grounding in code generation is still underexplored. Our work aims to fill this gap with a grounding-based pipeline for code generation and refinement, enhancing developer-model communication.

\section{Approach} \label{sec:approach}

\begin{figure*}
    \centering
    \includegraphics[width=\linewidth]{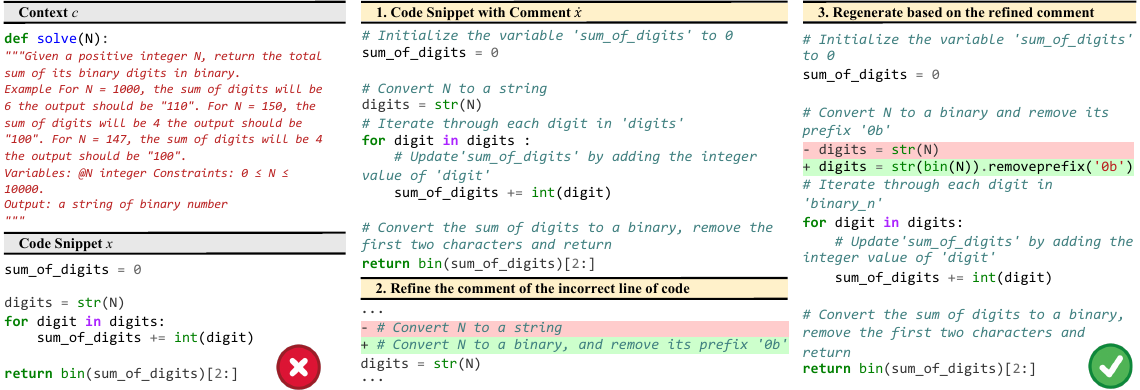}
    \caption{Refining a code snippet with our approach via comment editing.}
    \label{fig:example}
\end{figure*}

We formally define the task as follows: Given an initial context $c$ that comprises a natural language problem description and an initial code snippet $x$ produced by a code generation model, the task is to generate a refined code solution $\hat{x}$ that fixes the errors in $x$ and aligns the code more closely with the developer's intent as articulated in the problem description.

We introduce Programming with Interactive Grounding (\ping), which leverages inline code comments for bidirectional communication. \ping includes three steps: comment generation, human review, and code refinement. We elaborate on each step in the following sections.

\subsection{Comment Generation} \label{phase2}

\ping first parsed the initial code snippet $x$ into an Abstract Syntax Tree (AST) to segment the code snippet into individual statements. We opted for statement-level comment generation due to its fine-grained nature, which allows for a more accurate and detailed explanation of code behavior. This approach contrasts the previous approach that leverages coarser-grained code explanations or summaries, by providing targeted insights into specific statements of code.

We do not consider `import' and `definition' statements since they are trivial to explain. For compound statements such as `if', `for', `while', `try', and `with', we adopt a tailored approach for code segmentation. For such nodes with a conditional statement, the condition expression is first added to the statement list, followed by its body statement nodes. The statements in the `else' clause, if present, are handled afterward. For `try' nodes, the process directly navigates through its body nodes without additional processing. For `with' statements, the part of the statement excluding its `suite' is first added, after which is its suite part.

If the generated code snippet $x$ is not compilable, which prevents its transformation to AST, we resort to a simple strategy. In such cases, $x$ is directly segmented into individual lines for subsequent comment generation.

Given individual statements, a specialized comment generation model $M_C$ generates an inline comment $c$ for each of them. Figure \ref{fig:example} \circled{1} shows a code example with generated comments. When processing the for loop, \ping first generates an explanation for the loop condition. Then it proceeds to clarify the operations within the loop body.

To implement the comment generation model, we use CodeBERT~\cite{feng2020codebert} and fine-tune it with our \edit{finetuning} dataset. CodeBERT is a transformer-based encoder-only model. It is pre-trained on a diverse corpus of programming languages and natural language text. To develop the fine-tuning dataset, we adopted the approach outlined in the CodeBERT paper~\cite{feng2020codebert} and applied \edit{the} preprocessing \edit{steps from CodeXGLUE~\cite{lu2021codexglue}} to the CodeSearchNet~\cite{husain2019codesearchnet} dataset. \edit{The postprocessing process first cleans the data by (1) removing examples where the code cannot be parsed into an AST, (2) removing examples where the number of tokens in the document is fewer than 3 or greater than 256, (3) removing examples that contain special tokens (e.g., $\langle$img ...$\rangle$ or https://...), and (4) removing examples that are not written in English. Then, it extracts comments and their corresponding code to produce} a fine-tuning dataset with 280,652 code-to-comment pairs. We use Adam~\cite{kingma2014adam} to fine-tune the parameters. The learning rate was set to $5e^{-5}$, with a batch size of 32. We also set the maximum sequence length of input and inference as 256 and 128, respectively.

\subsection{Human Review} \label{phase3}

In the human review step, the code snippet with inline comments is presented to the developer. When users notice an error by reading the comments, they can directly modify the corresponding comment to describe the desired behavior. \edit{For example, in Figure \ref{fig:example} \circled{1}, the developer notices the statement in Line 4 is wrong based on the comment of that statement. Specifically, instead of converting the input \texttt{N} to a string, it should convert it to a binary first.} \edit{Fixing this bug requires changing this line of code to call three APIs: (1)  converting the number to binary via \code{bin}, (2)  transforming this binary to a string via \code{str}, and finally removing the string's prefix via \code{str.removeprefix}. The developer may not be familiar with the first and third APIs as they
are less commonly used. To manually fix it, the developer would typically need to search online and learn how to use these specific APIs first, which causes extra time and effort. Using \ping, the developer can modify the comment instead, as shown in  Figure~\ref{fig:example} \circled{2}. Then, \ping will regenerate the code based on the refined comment, which is more convenient for the developer.}

\subsection{Code Refinement} \label{phase4}

The original context $c$, the generated code up to the refined comment $\ddot{x}$, and the refined comments are concatenated to form a new, updated context $\hat{c}$. A code refinement model, $\hat{M}$, then processes $\hat{c}$ to regenerate code starting from the edited comment. This regeneration aims to address the identified error while maintaining the integrity of the correct parts.

To develop the code refinement model, we fine-tune the 6.7B version of DeepSeek Coder~\cite{guo2024deepseek}. DeepSeek Coder is a recently released code generation model with superior coding performance while being small enough to be fine-tuned on our GPUs. Instead of using the original code generation model, we decided to fine-tune it, since the original one is designed to generate code based on general, high-level task descriptions. However, in our code refinement step, the task is to generate a dedicated program statement based on a specific, detailed inline comment. Therefore, fine-tuning is necessary.

To construct a high-quality dataset for code refinement, we utilize the Stack dataset~\cite{Kocetkov2022TheStack}, which includes 190.73 GB of Python code files. We first follow the filtering methods of Codex~\cite{chen2021evaluating} and remove files with:
\begin{itemize}
    \item an average line length greater than 100 characters
    \item a maximum line length above 1,000 characters
    \item less than 25\% of the characters being alphanumeric
    \item keywords in the first few lines of the file indicating that the file was likely automatically generated
\end{itemize}

Then, we select code snippets that are sufficiently documented through comments. For each code file, we compute the comments-to-code ratio by counting the lines of comments and the lines of code. Finally, we set minimum and maximum thresholds for this ratio, selecting snippets within this range. After testing various thresholds, we established a MinRatio of 0.3 and a MaxRatio of 0.95 that optimally balances the number of code snippets against the richness of their comments.

We fine-tune the model with the next-token prediction objective, which is widely used in decoder-only models such as GPT-3. We use the Adam optimizer~\cite{kingma2014adam} with a learning rate of $2e^{-5}$ for updating the model weights. We use the Cross-Entropy~\cite{rubinstein1999cross} loss to optimize the model parameters. This loss compares the model's predicted probability distribution for the next token in a code or comment sequence against the ground-truth token. Mathematically, it is defined as:

\begin{equation}
    \mathcal{L} = -\sum_{i=1}^{N} y_i \log(\hat{y}_i)
\end{equation}

Here, $N$ denotes the dictionary size, $y_i$ is the ground-truth token, and $\hat{y}_i$ is the model's predicted token. We compared the performance of our fine-tuned model with the original model. Please refer to Section \ref{sec:experiments} for details.

\subsection{Implementation Details}

\begin{figure}
    \centering
    \includegraphics[width=\linewidth]{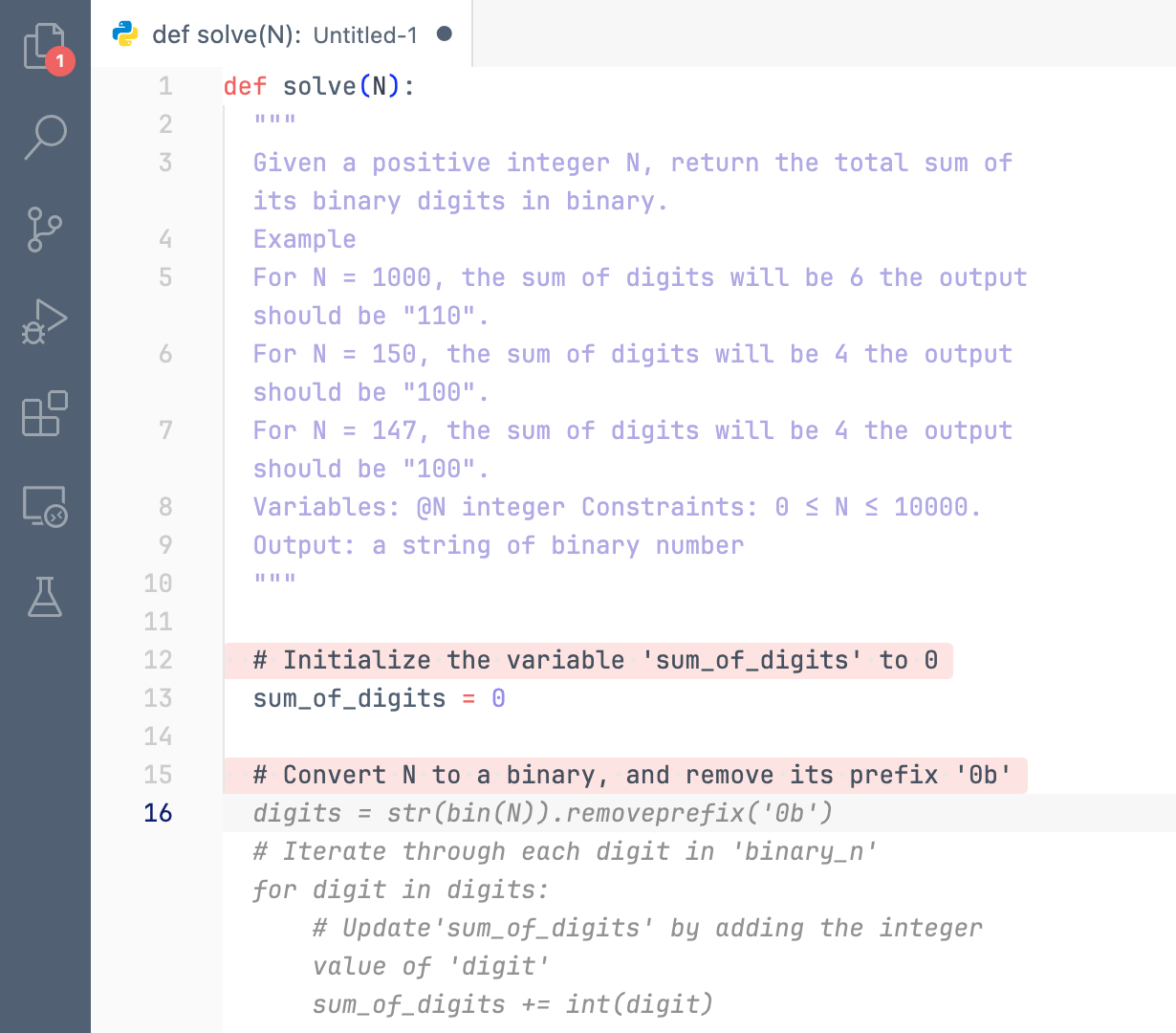}
    \caption{User Interface of \ping}
    \label{fig:ui}
\end{figure}

We have implemented \ping as an extension for Visual Studio Code (VSCode). Figure \ref{fig:ui} shows the user interface of this extension. The development of the \ping extension for VSCode involved writing 4,605 lines of code. The default models used in \ping are DeepSeek Coder for code generation, CodeBERT for comment generation, and the fine-tuned DeepSeek Coder model for code refinement.

\section{Simulated User Study} \label{sec:experiments}

In the evaluation, we strive to investigate to what extent {\ping} improves the accuracy of different kinds of code generation models in a broad range of programming tasks under different settings. Since {\ping} requires user feedback to guide the code refinement process, conducting such experiments at scale requires recruiting a large number of developers, which is costly and time-prohibitive to achieve in academia. Therefore, as a tradeoff, the first author acted as a simulated user and interacted with {\ping} in the experiments in this section. In other words, the first author manually inspected the code generated by {\ping} in each setting and edited the comments to provide feedback for code refinement (detailed in Section~\ref{sec:labeling}). With the simulated user, we investigate the following research questions:

\begin{itemize}
\item[$\bullet$] RQ1: How effectively can our interactive approach improve the code generation accuracy of different LLMs?
\item[$\bullet$] RQ2: How does our approach compare with other code generation approaches?
\item[$\bullet$] RQ3: How sensitive is our approach to different code comment generation models?
\item[$\bullet$] RQ4: To what extent can fine-tuning improve the code refinement accuracy?
\end{itemize}

In the end, we collected a very large interaction dataset, including 1224 code snippets generated by four different LLMs for 306 programming tasks, their inline comments generated by different comment generation models, comments edited by the simulated user, and code snippets generated by different code refinement models based on the edited comments. We have released this dataset to support the reproducibility of our experiments and also facilitate the development of new code refinement models for the research community. 

\subsection{Benchmarks}

We used two popular code generation benchmarks in this evaluation. \textbf{HumanEval}~\cite{chen2021evaluating} includes 164 hand-written Python programming tasks. \textbf{MBPP}~\cite{austin2021program} includes 974 crowd-sourced Python programming tasks. Since some tasks of MBPP have ambiguous task descriptions, the authors created a sanitized version with 427 tasks. For our evaluation, we randomly sampled 142 tasks from this sanitized version (about one-third) as analyzing all 427 tasks manually takes a lot of effort.

A recent study found that the original test cases from HumanEval do not sufficiently cover corner cases and thus cannot reliably assess the correctness of LLM-generated code~\cite{liu2023reliability}. The authors later released  \textbf{HumanEval+} and \textbf{MBPP+}, which extend HumanEval and MBPP with additional test cases. In our evaluation, we also measured model performance on HumanEval+ and MBPP+.

\subsection{User Feedback Collection}
\label{sec:labeling}

For the total of 306 programming tasks from HumanEval and MBPP, we experimented with four LLMs to generate the initial code snippets. We selected these models since they are well-known code generation models with different performance levels. We explain each of them below.

\textbf{CodeGen}~\cite{nijkamp2022codegen} leverages multi-task learning on textual and programming language tasks, CodeGen can effectively generate syntactically valid code while capturing natural language meaning. For our evaluation, we used its 16B version.

\textbf{InCoder}~\cite{fried2022incoder} utilizes a retrieve-and-edit approach in code generation. Given a text description, it first retrieves relevant code snippets and then edits them to match the input. Its hybrid pointer-generator network allows both copying and generating tokens.
For our evaluation, we used its 6B version.

\textbf{Code-davinci-002} is OpenAI's GPT-3-based code generation model. It generalizes well across programming languages without task-specific training. As a close-sourced model, its model size is undisclosed. We used the API provided by OpenAI to access this model and test it in our experiments.

\textbf{DeepSeek Coder}~\cite{guo2024deepseek} is a new series of code language models that show superior coding performance. These models are pre-trained with a large project-level code corpus with both the next token prediction and the fill-in-the-middle objectives. For our evaluation, we used its 6.7B version.

\edit{For each task, we analyzed the top one program generated by each model. This resulted in 1,224 code snippets. 738 of them fail to pass the test cases. Table~\ref{tab:models} shows the distribution of these 738 incorrect solutions across these models. Two annotators manually inspected these incorrect code solutions and edited the comments to refine the code. One of the annotators is the first author, a graduate student with over nine years of programming experience, including three years of industry experience. The other annotator, who is not the co-author of this paper, is a graduate student invited externally with five years of programming experience. We describe the detailed annotation procedure below.}

\edit{The first step is to identify the buggy statements in each code snippet by examining the comments. The two annotators first had a 30-minute session to get familiar with the annotation task and go over 10 code snippets together to practice. Subsequently, they split the remaining 728 code snippets into halves and each independently inspected 364 code snippets. During the inspection and fault localization process, they discussed in case of uncertainty. After finishing their first round of inspection, they exchanged to check each other's results. The initial agreement level between the two annotators in this step is 0.85 in terms of Cohen’s Kappa, indicating a perfect agreement level~\cite{mchugh2012interrater}. This agreement level makes sense since, in most cases, it is obvious which line of code is incorrect. The annotators then discussed and resolved all disagreements.}

\edit{The second step is to edit the comment to reflect the expected behavior of a buggy statement for \ping to refine the code. Like the previous step, they first practiced together on 10 snippets and then independently edited comments in half of the remaining snippets. Upon completion, they inspected the edited comments with each other and checked whether they agreed on the edited comments. The initial Cohen's Kappa is 0.72, indicating a substantial agreement level~\cite{mchugh2012interrater}. Compared with the previous step, the agreement level is a bit lower since, in some cases, the annotators disagreed on the clarity and specificity of edited comments. For example, one annotator found another annotator's comment edit ambiguous, too trivial, or even confusing. The annotators discussed each comment edit they initially disagreed on and came up with a final edit that both of them found appropriate.}

\edit{They then ran \ping to refine the code with the edited comment for each code snippet. If the refined code still failed the test cases, the two annotators repeated the previous steps to provide feedback to the refined code. Due to the enormous manual effort in this process, the annotators only provided up to three rounds of feedback to the 251 incorrect solutions from the HumanEval dataset.}

\edit{After this annotation process}, we logged all the generated code in each iteration and code comments before and after editing to form the interaction dataset described at the beginning of this section. We reported the performance gains of each round of feedback in Table~\ref{tab:progressive}. 

\begin{table}
\centering
\caption{Comparison of code generation models evaluated}
\label{tab:models}
\resizebox{.48\textwidth}{!}{
\begin{tabular}{@{}llrrr@{}}
\toprule
\textbf{}                              & \multicolumn{1}{c}{\multirow{2}{*}{\textbf{\begin{tabular}[c]{@{}c@{}}Model\\ Size\end{tabular}}}} & \multicolumn{2}{c}{\textbf{Pass@$1$}}                                        & \multicolumn{1}{c}{\multirow{2}{*}{\textbf{\begin{tabular}[c]{@{}c@{}}Incorrect \\ Solution \#\end{tabular}}}} \\
                                       & \multicolumn{1}{c}{}                                                                               & \multicolumn{1}{c}{\textbf{HumanEval}} & \multicolumn{1}{c}{\textbf{MBPP}} & \multicolumn{1}{c}{}                                                                                           \\
\midrule
\textsc{InCoder}          & 6B          & 16.5          & 19.7          & 262 \\
\textsc{CodeGen-Mono}     & 16B         & 29.9          & 35.2          & 212 \\
\textsc{code-davinci-002} & Undisclosed & 46.3          & 58.5          & 165 \\
\textsc{DeepSeek-Coder}   & 6.7B        & 74.4          & 73.2          & 99 \\
\bottomrule
\end{tabular}
}
\end{table}

\subsection{Evaluation Metrics}

Following the Codex paper\cite{chen2021evaluating}, we measured each model's performance using \pass{k}. We set $k$ to 1 and only considered the top 1 program generated by each model in our experiments.

\subsection{Comparison Baselines}
Since RQ1 only aims to measure the performance gain achieved on different LLMs, we compared the \pass{1} of each model with and without the augmentation of \ping. We also compared the \pass{1} of each model after each feedback iteration and reported the improvement trend over iterations.

\begin{table*}
\centering
\caption{Pass@$1$ ($\%$) on the HumanEval and HumanEval+ datasets}
\label{tab:humaneval}
\begin{tabular}{lrllrll} 
\toprule
\multicolumn{1}{c}{} & \multicolumn{3}{c}{\textbf{HumanEval}} & \multicolumn{3}{c}{\textbf{HumanEval+}}  \\ 
\multicolumn{1}{c}{}     & \multicolumn{1}{c}{\textbf{Original}} & \multicolumn{1}{c}{\textbf{\ping}} & \multicolumn{1}{c}{\textbf{\ping w/o Finetuning}} & \multicolumn{1}{c}{\textbf{Original}} & \multicolumn{1}{c}{\textbf{\ping}} & \multicolumn{1}{c}{\textbf{\ping w/o Finetuning}}  \\ 
\midrule
\incoderb & 16.5 & \edit{29.9}~\improve{13.4} & \edit{26.2}~\improve{9.7} & 12.2 & \edit{25.6}~\improve{13.4} & \edit{23.8}~\improve{11.6} \\
\codegenb & 29.9 & \edit{43.9}~\improve{14.0} & \edit{41.5}~\improve{11.6} & 28.0 & \edit{41.5}~\improve{13.5} & \edit{40.2}~\improve{12.2}  \\
\davincitwo & 46.3 & \edit{63.4}~\improve{17.1} & \edit{63.4}~\improve{17.1} & 42.1 & \edit{59.1}~\improve{17.0} & \edit{59.1}~\improve{17.0}  \\
\deepseek & 74.4 & \edit{79.9}~\improve{5.5} & \edit{78.7}~\improve{4.3} & 70.7 & \edit{76.2}~\improve{5.5} & \edit{74.4}~\improve{3.7}  \\
\bottomrule
\end{tabular}
\end{table*}

\begin{table*}
\centering
\caption{Pass@$1$ ($\%$) on the MBPP and MBPP+ datasets}
\label{tab:mbpp}
\begin{tabular}{lrllrll} 
\toprule
\multicolumn{1}{c}{} & \multicolumn{3}{c}{\textbf{MBPP}} & \multicolumn{3}{c}{\textbf{MBPP+}}  \\ 
\multicolumn{1}{c}{}     & \multicolumn{1}{c}{\textbf{Original}} & \multicolumn{1}{c}{\textbf{\ping}} & \multicolumn{1}{c}{\textbf{\ping w/o Finetuning}} & \multicolumn{1}{c}{\textbf{Original}} & \multicolumn{1}{c}{\textbf{\ping}} & \multicolumn{1}{c}{\textbf{\ping w/o Finetuning}}  \\ 
\midrule
\incoderb & 19.7 & \edit{30.3}~\improve{10.6} & \edit{27.5}~\improve{7.8} & 16.9 & \edit{26.8}~\improve{9.9} & \edit{24.6}~\improve{7.7} \\
\codegenb & 35.2 & \edit{45.8}~\improve{10.6} & \edit{45.1}~\improve{9.9} & 33.1 & \edit{44.4}~\improve{11.3} & \edit{43.0}~\improve{9.9}  \\
\davincitwo & 58.5 & \edit{69.0}~\improve{10.5} & \edit{69.0}~\improve{10.5} & 50.7 & \edit{63.4}~\improve{12.7} & \edit{63.4}~\improve{12.7}  \\
\deepseek & 73.2 & \edit{78.2}~\improve{5.0} & \edit{76.8}~\improve{3.6} & 64.1 & \edit{69.7}~\improve{5.6} & \edit{67.6}~\improve{3.5}  \\
\bottomrule
\end{tabular}
\end{table*}

RQ2 compares {\ping} with other code generation and refinement techniques. Thus, we selected eight state-of-the-art methods as the baselines. \edit{To ensure comparison consistency, we use GPT-3.5 with the default temperature (1.0) and top\_p (1.0) as the base model for \ping and all baselines in our experiments. For iterative methods such as Self-Edit~\cite{zhang2023self} and Self-Debugging~\cite{chen2023teaching}, we set the iteration upper bound to three, since the performance of these methods often saturates after two or three iterations based on the experiments in their papers. We describe each method below.}

\begin{itemize}
    \item \textbf{ReAct}~\cite{yao2022react}. ReAct prompts an LLM to generate both a reasoning trace and an action plan in an interleaved manner. This allows the model to perform dynamic reasoning to adjust the action plan.
    \item \textbf{ToT}~\cite{yao2023tree}. ToT structures potential reasoning paths in a tree-like manner, which goes beyond the linear reasoning of traditional Chain-of-Thought (CoT) prompting to enable the exploration of multiple reasoning pathways.
    \item \textbf{RAP}~\cite{wang2023rap}. RAP enhances LLM's problem-solving by using them as both reasoning agents and world models to generate actionable plans and conduct complex reasoning, which predict environmental states and simulate action outcomes for more effective problem-solving.
    \item \textbf{Self-Edit}~\cite{zhang2023self}. Self-Edit improves code generation accuracy by executing the generated code, analyzing results, and guiding the fault-aware code editor to correct errors in a generate-and-edit cycle.
    \item \textbf{Self-Planning}~\cite{jiang2023self}. Self-Planning employs a two-phase (planning and implementation) approach in LLMs to enhance LLMs' understanding and handling of complex code generation tasks.
    \item \textbf{Self-Debugging}~\cite{chen2023teaching}. Self-Debugging empowers LLMs to perform rubber duck debugging on their own generated code via few-shot demonstrations, natural language code explanation, and execution analysis.
    \item \textbf{SCoT}~\cite{li2023enabling}. SCoT improves traditional Chain-of-Thought (CoT) prompting by incorporating the program structures to obtain structured CoTs, which leads to more organized and efficient code generation.
    \item \textbf{CodeChain}~\cite{le2023codechain}. CodeChain prompts LLMs to generate modular code and refine it via self-revisions. It boosts accuracy by extracting, clustering, and reusing code sub-modules, emulating expert programming practices.
\end{itemize}

For RQ3, we evaluated how \ping performs with a different comment generation model,  Seq2Seq~\cite{sutskever2014sequence}. Prior research has validated Seq2Seq's effectiveness in generating inline comments~\cite{huang2023comparative}. We compared the \pass{1} rates when using the proposed CodeBERT model and the Seq2Seq model.

\section{Experiment Results}

\subsection{RQ1: Effectiveness on Different Code Generation Models}

\begin{table}
\centering
\caption{Pass@$1$ rates across multiple iterations on HumanEval}
\label{tab:progressive}
\resizebox{.48\textwidth}{!}{
\begin{tabular}{@{}lrlll@{}}
\toprule
\multirow{2}{*}{} & \multicolumn{4}{c}{\textbf{HumanEval}}                         \\ 
                  & \textbf{Original} & \textbf{1 iteration}  & \textbf{2 iterations} & \textbf{3 iterations} \\
\midrule
\incoderb           & 16.5     & \edit{26.2}~\improve{9.7} & \edit{31.1}~\improve{4.9}  & \edit{34.1}~\improve{3.0}  \\
\codegenb           & 29.9     & \edit{41.5}~\improve{11.6} & \edit{45.7}~\improve{4.2}  & \edit{48.2}~\improve{2.5}  \\
\davincitwo         & 46.3     & \edit{63.4}~\improve{17.1} & \edit{67.7}~\improve{4.3}  & \edit{70.1}~\improve{2.4}  \\
\deepseek           & 74.4     & \edit{78.7}~\improve{4.3}  & \edit{81.7}~\improve{3.0}  & \edit{83.5}~\improve{1.8}  \\
\bottomrule
\end{tabular}}
\end{table}

Table \ref{tab:humaneval} shows the \pass{1} results after a single refinement iteration of \ping across different LLMs on both HumanEval and its more rigorous version, HumanEval+. Similarly, Table \ref{tab:mbpp} shows the \pass{1} results after a single refinement iteration of \ping on MBPP and its more rigorous version, MBPP+.

Results shown in Table \ref{tab:humaneval} reveal significant improvements in code generation accuracy through just one iteration of refinement. For instance, the code-davinci-002 model shows a remarkable improvement on the HumanEval dataset, with the \pass{1} rate improved from 46.3\% to \edit{63.4}\%. This enhancement highlights the significant influence of focused, iterative feedback in correcting model misconceptions.

Similar patterns of improvement were discernible across other models, including InCoder and CodeGen. Notably, InCoder's performance on the HumanEval dataset ascended from 16.5\% to \edit{29.9}\%, while CodeGen's accuracy experienced a boost from 35.2\% to \edit{45.8}\% on the MBPP benchmark. These improvements again confirm the effectiveness of incorporating human insights into the code generation process.

The overarching findings from our investigation affirm the critical role of human feedback in guiding LLMs toward synthesizing more functionally accurate programs. The consistent improvements across different models further validate the hypothesis that collaborative interaction is key to effectively addressing the challenges of code generation. This collaborative paradigm facilitates a model's ability to dynamically adjust its outputs based on constructive feedback, thereby incrementally moving closer to generating error-free, functional code.

We also conducted experiments to explore how extending the interactive refinement process beyond a single iteration impacts the improvements in code generation. Table \ref{tab:progressive} demonstrates a consistent pattern of improvement across all evaluated models, with each additional iteration yielding positive gains, albeit at a diminishing rate. For instance, the InCoder model shows a continuous gain from a baseline of 16.5\% to \edit{34.1}\% after three iterations, marking a total improvement of \edit{17.6}\%. Similarly, the CodeGen model's performance ascends progressively, culminating in a \edit{48.2}\% \pass{1} rate, while the code-davinci-002 model reaches a peak of \edit{70.1}\%.

In summary, our analysis of both single and multiple iterations of interactive refinement reveals a clear trajectory of improvement in code generation accuracy. The gains were consistent across diverse model architectures, showing collaboration is vital for complex code generation. The findings highlight the efficacy of human feedback in guiding LLMs toward higher accuracy levels, affirming the value of interactive approaches in code generation. 

\finding{\ping significantly improves the code generation capability of four different LLMs on multiple benchmarks, demonstrating the effectiveness of comment-level code refinement.}

\subsection{RQ2: Comparison to Other Prompting Approaches}

\begin{table}
\centering
\caption{Comparison of \ping with other code generation approaches on HumanEval and MBPP benchmarks}
\label{tab:rq1}
\begin{tabular}{lrrrr}
\toprule
& \multicolumn{2}{c}{\textbf{HumanEval}} & \multicolumn{2}{c}{\textbf{MBPP}} \\
& \textbf{Pass@$\textbf{1}$} & \textbf{Time (sec)} & \textbf{Pass@$\textbf{1}$} & \textbf{Time (sec)}    \\
\midrule
ReAct~\cite{yao2022react}              & 56.7                & \edit{50.8}                & 66.9              & \edit{42.6}             \\
ToT~\cite{yao2023tree}                 & 54.3                & \edit{62.4}                & 65.5              & \edit{51.3}             \\
RAP~\cite{wang2023rap}                 & 62.8                & \edit{59.6}                & 70.4              & \edit{50.9}             \\
Self-Edit~\cite{zhang2023self}         & 62.2                & \edit{40.1}               & 56.3              & \edit{35.7}             \\
Self-Planning~\cite{jiang2023self}     & 65.2                & \edit{37.8}                & 58.5              & \edit{34.5}             \\
Self-Debugging~\cite{chen2023teaching} & 61.6                & \edit{42.7}                & 59.9              & \edit{36.1}            \\
SCOT~\cite{li2023enabling}             & 61.0                & \edit{48.2}                & 46.5              & \edit{44.2}             \\
CodeChain~\cite{le2023codechain}       & 62.8                & \edit{64.5}                & 59.2              & \edit{58.7}             \\
\textbf{\ping}            & \textbf{65.9}       & \edit{35.2}       & \textbf{71.1}     & \edit{29.3} \\
\bottomrule
\end{tabular}
\end{table}

Table \ref{tab:rq1} shows the \pass{1} results of different code generation and refinement methods on the HumanEval benchmark. \edit{Note that the results on \ping in this table are only based on one iteration of human feedback.} The results highlight the performance of our approach, \ping, compared to other leading LLM-based approaches in code generation.

Table \ref{tab:rq1} demonstrates that \ping not only performs competitively but also surpasses other approaches on both HumanEval and MBPP benchmarks. It proves to be a highly effective approach for addressing the complexities of code generation and refinement with \pass{1} of \edit{65.9} and \edit{71.1}, respectively.

On the HumanEval benchmark, \ping's leading score of \edit{65.9} marks a notable advancement in code generation, exceeding Self-Planning, the closest competitor. This performance illustrates \ping's effectiveness in producing correct code from natural language task descriptions. Similarly, on the MBPP benchmark, \ping outperforms RAP, demonstrating superior refinement capabilities by integrating human feedback. This success is largely due to \ping's innovative use of inline comments to align generated code with user intent for more precise code refinement.

\finding{\ping outperformed state-of-the-art code generation and refinement methods, demonstrating its effectiveness in leveraging human feedback for more accurate and contextualized code refinement.}

\subsection{RQ3: Sensitivity to Code Comment Generation Models}

\begin{table}
\centering
\caption{Pass@$1$ with different comment generation models on the HumanEval dataset}
\label{tab:comment-gen}
\begin{tabular}{@{}lrll@{}}
\toprule
 & \multicolumn{1}{c}{\textbf{Original}} & \multicolumn{1}{c}{\textbf{Seq2Seq}} & \multicolumn{1}{c}{\textbf{CodeBERT}} \\ \midrule
\incoderb                 & 16.5                         & \edit{23.8}~\improve{7.3}                        & \edit{26.2}~\improve{9.7}                         \\
\codegenb                    & 29.9                         & \edit{39.0}~\improve{9.1}                        & \edit{41.5}~\improve{11.6}                       \\
\davincitwo           & 46.3                         & \edit{60.4}~\improve{14.1}                        & \edit{63.4}~\improve{17.1}                         \\ \bottomrule
\end{tabular}
\end{table}

Table \ref{tab:comment-gen} compares the \pass{1} rates of \ping on different LLMs when CodeBERT vs.~Seq2Seq as the code comment generation model. The results reveal an advantage in favor of using CodeBERT over Seq2Seq across all LLMs.
However, the impact of using different code comment generation models is not significant for all three LLMs---around 2\% to 3\% differences in \pass{1}. This implies that while using a better comment generation model can contribute to the refinement process, the advantage of \ping is more pronounced by leveraging human feedback to refine code comments. Even with the suboptimal Seq2Seq model, the \pass{1} improvements over the original LLMs are around 7\% to 14\%. 

\edit{In addition to measuring the impact of the code comment model on \pass{1}, we further measured the accuracy of the CodeBERT model. Due to the lack of ground-truth comments, the same two annotators from Section \ref{sec:labeling} sampled 385 code-comment pairs from the simulation dataset and categorized the comments into three accuracy levels: {\em fully accurate}, {\em largely accurate but with missing information}, and {\em contain wrong information}. This sample size is statistically significant with a 95\% confidence level and a margin of error of 5\%. The annotators followed a similar data annotation process as in Section~\ref{sec:labeling} to first have a 1-hour session to go over 25 code-comment pairs together. Then, they independently annotated half of the remaining 360 pairs and exchanged their annotations for validation. The initial agreement between the two annotators is 0.76 in terms of Cohen’s Kappa, indicating a substantial agreement level~\cite{mchugh2012interrater}. Then, they discussed the disagreements to achieve a consensus.} 

\edit{Among 385 code-comment pairs, the majority (72\%) of comments generated by \ping are fully correct. 12\% of the comments are largely correct but with missing information, while 16\% contain wrong information. Given that \ping only regenerates the code for user-revised comments, missing or wrong information in other unchanged comments has little impact on the regenerated code. Yet we acknowledge that they may confuse or distract developers, since developers may spend extra time scrutinizing the code and comments to figure out whether the missing information is because of incorrect code or simply a comment generation error.}

\finding{\ping's performance is not significantly impacted by the choice of comment generation model, illustrating its robustness and the primary value of iterative human feedback in enhancing code generation accuracy.}

\subsection{RQ4: The Impact of Finetuning}

RQ4 investigates how fine-tuning the code refinement model affects the code generation accuracy of \ping. Column \textsf{\ping} and Column \textsf{\ping w/o Finetuning} in Table \ref{tab:humaneval} and \ref{tab:mbpp} compare the results of \ping with a fine-tuned code refinement model against its performance without fine-tuning. 

Both the InCoder and CodeGen models show notable improvements in \pass{1} rates on both datasets with fine-tuning. InCoder's performance increased by \edit{3.7}\% on HumanEval and \edit{2.8}\% on MBPP, while CodeGen saw a \edit{2.4}\% gain on HumanEval. These results highlight the effectiveness of fine-tuning in enhancing code refinement accuracy and understanding of user intent.

\finding{Fine-tuning code generation models enhances code refinement accuracy, with our \ping approach yielding marked improvements over baselines and fine-tuning offering additional performance boosts.}

\section{User Study} \label{sec:user-study}

To evaluate the real-world utility and usability of our grounding-based approach, we utilize our integrated extension for Visual Studio Code to conduct a within-subjects user study with 12 participants, including both college students and professional programmers. This study aims to investigate the following research question:

\begin{itemize}
\item \textbf{RQ5:} How useful is our interactive approach
to real programmers in practice?
\end{itemize}

\subsection{Participants}
We recruited 12 participants (10 males, 2 females) from a diverse background, comprising 2 graduate students, and 10 professional developers. They were recruited by emailing student mailing lists at an R1 university and reaching out to experienced software developers through our personal network. 2 had 1-2 years of Python experience (early intermediate), 5 had 3-5 years of experience (late intermediate), and 5 participants had over 5 years of Python experience (expert). This choice allows us to assess how well our grounding-based approach performs across a spectrum of Python expertise.

\subsection{Task}

Our task selection was inspired by the programming tasks used by Xu et al.~\cite{xu2021ide} and Vaithilingam et al.~\cite{vaithilingam2022expectation}. We first categorized the TranX Developer Study tasks~\cite{xu2021ide} and tasks from the DS-1000 benchmark~\cite{lai2023ds} into different common programming task types. We then employed stratified random sampling from the two datasets. This process resulted in a pool of 6 code generation tasks with different types, including File I/O, OS, Web Scraping, Web Server \& Client, Data Analysis, and Data Visualization. The ground-truth code for those tasks ranges from 15 LOC to 50 LOC.

\subsection{Comparison Baselines}

In addition to \ping, we used GitHub Copilot~\cite{copilot2023} and Multi-Turn Program Synthesis (Multi-Turn for short from henceforth)~\cite{nijkamp2022codegen} as comparison baselines. GitHub Copilot, powered by Codex, is an AI assistant that suggests code completions but lacks explicit repair feedback loops. Multi-Turn is a paradigm that decomposes code generation into steps, where the model generates subprograms in response to natural language instructions. Here, users can only accept the generated code and provide further instructions for subsequent steps. The individually generated subprograms are concatenated into a single, complete program as the final generated program. GitHub Copilot offers continuous code generation without direct feedback mechanisms, while Multi-Turn enables task decomposition and multi-turn interaction without any chances for refinement.
We used the VSCode extension of GitHub Copilot for our studies. For Multi-Turn, since it originally lacked a VSCode extension, we developed one with a similar user interface to \ping's to ensure a fair comparison.

\subsection{Procedure}

Our study employed a within-subjects design to directly compare \ping with two comparison baselines. The whole study lasted about 100-115 minutes.

Participants first received an overview of the procedure, completed a consent form, and filled out a pre-study questionnaire that collected their background information. Then, each study was divided into three sessions, with each focusing on a specific code generation tool. 

During these sessions, participants were asked to use the designated tool to tackle the assigned programming task within 20 minutes. \edit{To minimize learning effects, the assignment orders of both tasks and tools are counterbalanced across participants to ensure that each task-tool pair has the same number of trials}. \edit{In each study, a participant was required to use all three different tools on three different tasks, with the ordering of the three task-tool pairs randomized.} 

Each session began with a tutorial video about the particular tool in focus. A subsequent 5-minute practice period allowed participants to get familiar with the tool before delving into the actual tasks. For each task, participants were asked to study its task description and use the tool to generate the code first. Participants could review the code and then use the tool's interactive features to refine it until they deemed the code to be correct. \edit{Participants are allowed to adopt any approaches to verify the correctness of their current programs, including writing unit tests.} If participants found a task too challenging, they could choose to skip it.

After each session, participants filled out a post-task questionnaire to assess confidence in their final code, perceived success rates, and five different cognitive load questions from NASA TLX~\cite{hart1988development}. After wrapping up all sessions, participants completed a final survey to compare their experiences of using these three tools. We record all the screen activities for playback to aid in discussing observed behaviors. All the systems logged interaction events, including code editing and comment editing, to analyze participants' behavior patterns.

\subsection{User Performance Results}

\begin{figure}[tbp]
    \centering
    \includegraphics[width=\linewidth]{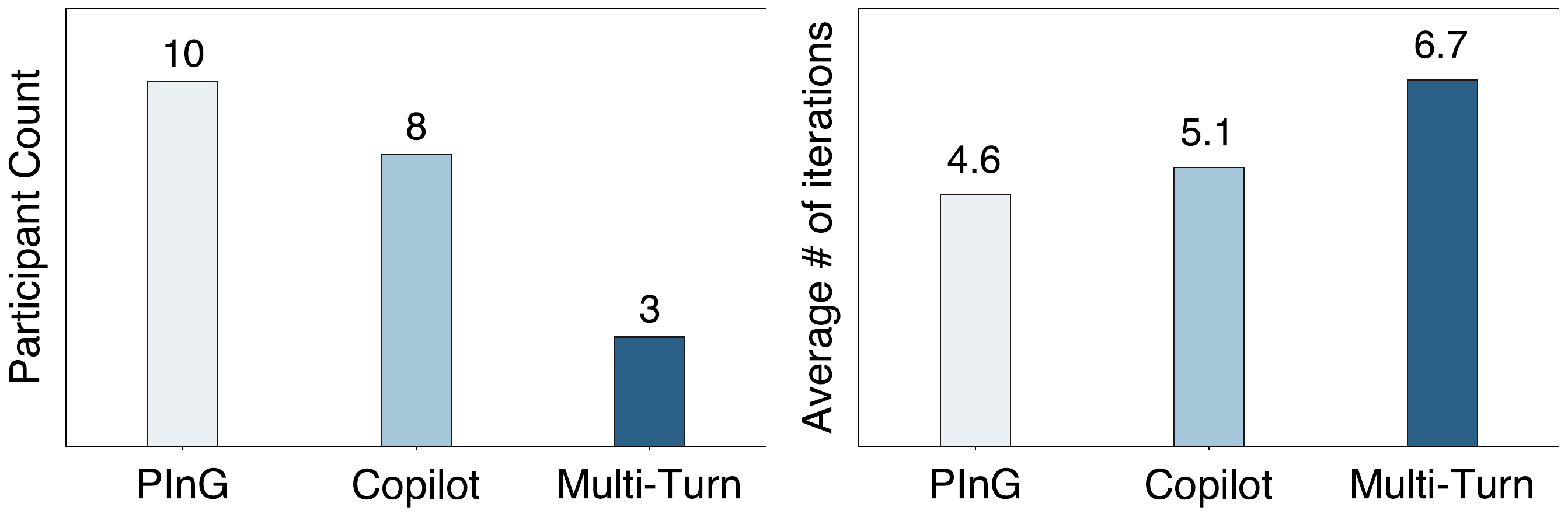} 
    \caption{\edit{The number of participants whose final code was functionally correct and the average number of iterations these participants took.}}
    \label{fig:user_completion}
\end{figure}

\begin{table}[tbp]
    \centering
    \caption{The average task completion time (in Minutes)}
    {
    \begin{tabular}{c|ccc}
    \toprule
    {Task Type} & {\textit{PInG}} & {\textit{GitHub Copilot}} & {\textit{Multi-Turn}}  \\
    \midrule
    File I/O & 9.6 & 10.8 & 15.2 \\
    OS & 13.7 & 16.2 & 18.8 \\
    Web Scraping & 16.3 & 19.4 & 20 \\
    Web Server \& Client & 19.7 & 20 & 20 \\
    Data Analysis & 11.5 & 13.6 & 20 \\
    Data Visualization & 17.1 & 18.2 & 20 \\
    \midrule
    Average & 14.65 & 16.37 & 19 \\
    \bottomrule
    \end{tabular}}
\label{tab:user-study-settings}
\end{table}

As shown in Figure \ref{fig:user_completion}, 10 of 12 participants successfully solved the assigned programming task using \ping within the given time. In comparison, 8 participants solved the given task with GitHub Copilot, and 3 with Multi-Turn. Compared to GitHub Copilot and Multi-Turn, \ping improved the success rate by 16.7\% and 58.3\%, respectively. It showcases the effectiveness of solving programming tasks by using inline comments to guide code generation.

Table \ref{tab:user-study-settings} compares task completion times across different tools. On average, participants finished their tasks 10.5\% and 22.9\% faster with \ping when compared to GitHub Copilot and Multi-Turn, respectively. \edit{We further analyzed the average number of iterations participants tried to complete the given programming task with the assigned tool. Figure \ref{fig:user_completion} shows these results. On average, participants using \ping required fewer iterations (4.6) compared to GitHub Copilot (5.1) and Multi-Turn (6.7) to complete a programming task. This highlights how leveraging comments as a communication vehicle between users and LLMs improves the efficiency of identifying and resolving errors in LLM-generated code.}

\edit{Although our study did not prohibit participants from writing unit tests, we observed that most participants chose not to write tests during the task. Instead, they relied on the inline comments generated by \ping to inspect and understand code functionality quickly and effectively. At the end of each task, some participants opted to write one or two unit tests to double-check the correctness of their final code. This reflects a practical balance between immediate, comment-driven insights and traditional unit testing for comprehensive validation.}

\subsection{User Confidence and Cognitive Overhead Result}

\begin{figure}[t]
    \centering
    \includegraphics[width=\columnwidth]{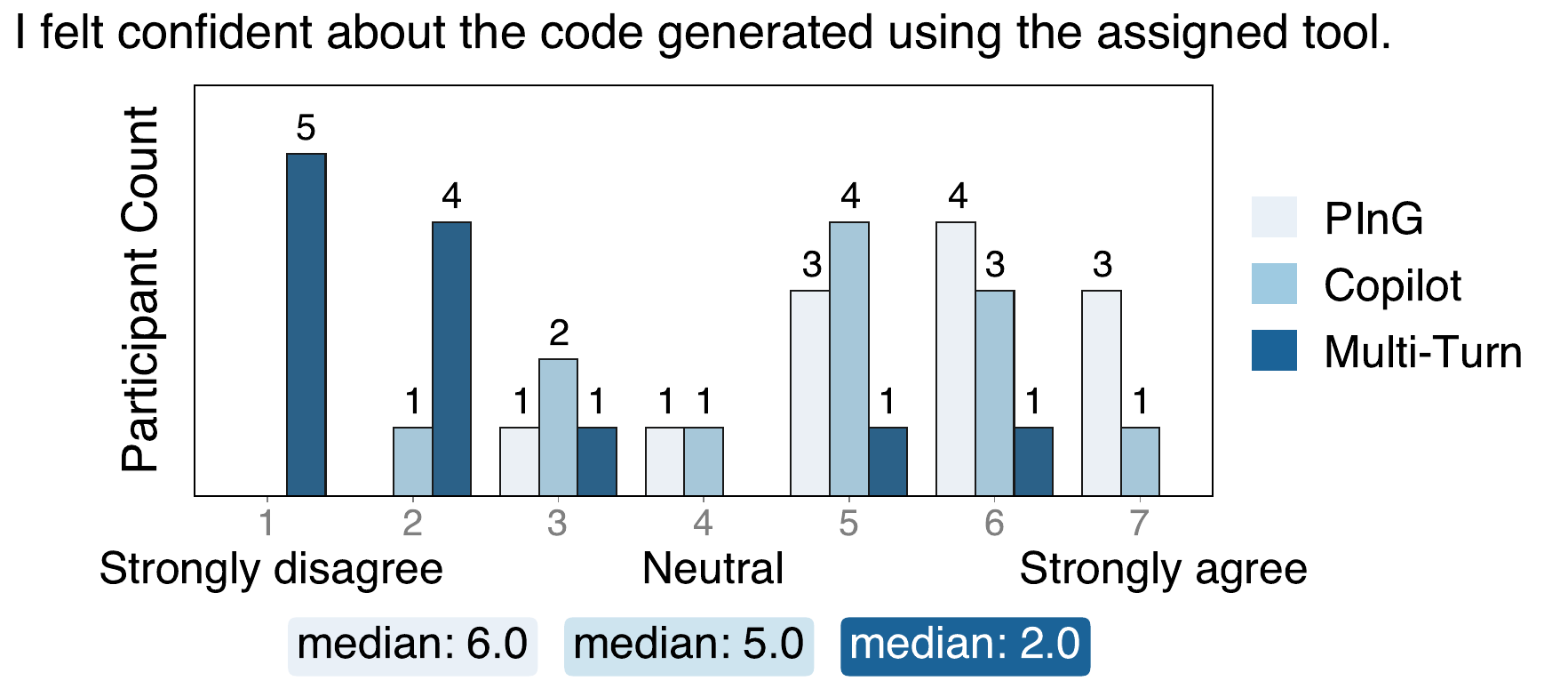}
    \caption{The distribution of participants’ confidence in the final code generated using the assigned tools.}
    \label{fig:user_confidence}
\end{figure}

In the post-study survey, \edit{participants self-reported their confidence in solving the given programming task on a 7-point scale (1—very low confidence, 7—very high confidence).} Notably, 9 of 12 participants agreed or strongly agreed that \ping helped generate code that aligned with their intent, compared to 7 participants for GitHub Copilot and just 1 for Multi-Turn. Furthermore, \ping significantly enhances developers’ confidence in auto-generated patches. As shown in Figure \ref{fig:user_confidence}, 7 participants agreed or strongly agreed that they felt confident about the generated code when using \ping. In contrast, only 4 and 1 participant felt so when using GitHub Copilot and Multi-Turn, respectively. This can be largely attributed to \ping's interactive refinement process, which empowers users to iteratively refine the generated code through inline comments, offering a more transparent and controlled development experience. This stark contrast in user confidence levels underscores the value of \ping's approach, emphasizing the importance of a feedback loop in fostering trust and satisfaction with AI-generated code.

\begin{figure}[t]
    \centering
    \includegraphics[width=\columnwidth]{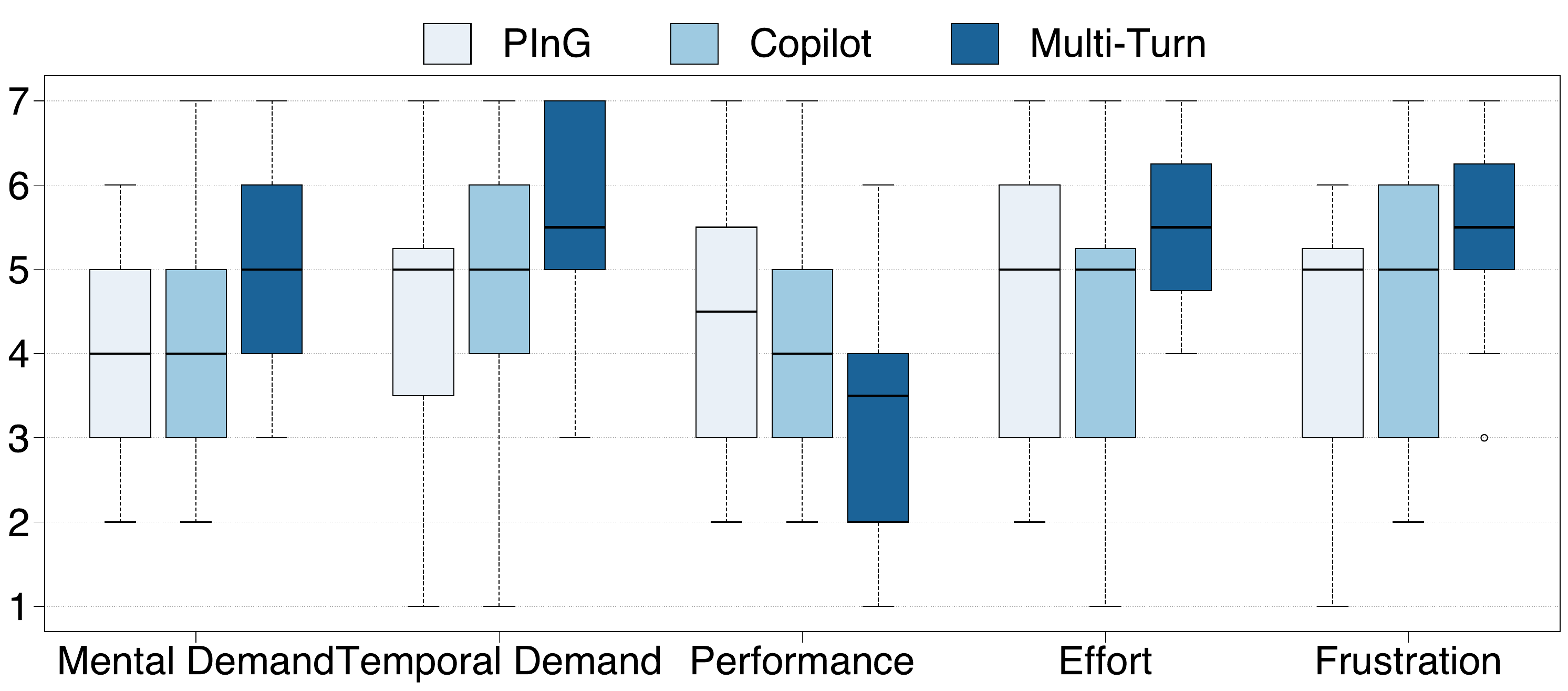}
    \caption{\edit{The distribution of participants’ cognitive load when solving the given programming tasks with the assigned tools.}}
    \label{fig:cognitive_load}
\end{figure}

\edit{As shown in Figure \ref{fig:cognitive_load}, solving programming tasks with \ping generally leads to a better user experience across all five cognitive load metrics compared to using Multi-Turn. Three of the five metrics show statistically significant differences between using \ping and Multi-Turn based on the Wilcoxon signed-rank test (\textit{Performance}: $p=0.047$, \textit{Effort}: $p=0.038$, and \textit{Frustration}: $p=0.022$). This shows the benefits of using comments as the communication vehicle between users and LLMs. Compared to Copilot, \ping slightly reduces the programming workload in terms of \textit{Mental Demand}, \textit{Temporal Demand}, \textit{Performance}, and \textit{Frustration}. However, these differences are not statistically significant.}

Additionally, the user study shows our interactive approach assists real-world programming in multiple ways. For simple tasks, 67\% of the participants directly used the code with minor edits. For complex tasks, while extensive changes were required, over 83\% agreed the generated code gave useful structural scaffolds to build on. 75\% of them reported that the inline comments enabled quickly identifying and fixing erroneous portions rather than debugging from scratch. On average, from 2-3 iterations, participants felt the code matched specifications sufficiently to serve as a quality starting point.

\subsection{Impact of Task Difficulty Level}

\edit{We categorized the programming tasks into three levels: easy, medium, and hard. Each level contains two tasks, which are classified based on the lines of code (LOC) in the ground truth code for each task. Tasks \textsf{File I/O} (16 LOC) and \textsf{Data Analysis} (22 LOC) are categorized as easy tasks. Tasks \textsf{OS} (27 LOC) and \textsf{Web Scraping} (37 LOC) are categorized as medium. Tasks \textsf{Web Server \& Client} (46 LOC) and \textsf{Data Visualization} (51 LOC) are categorized as hard.}

\edit{Table \ref{tab:difficulty_success_rate} shows the success rates for each difficulty level of programming tasks. We observe that participants using \ping outperform those using the other two tools in medium and hard tasks. In easy tasks, \ping and GitHub Copilot demonstrate equivalent success rates, both outperforming Multi-Turn. Table \ref{tab:difficulty_completion_time} shows the average task completion time for each difficulty level. \ping consistently saves time compared with GitHub Copilot and Multi-Turn across all difficulty levels. }

\subsection{Impact of User's Programming Expertise}

The impact of programming expertise on solving programming tasks with code generation tools is noteworthy. \edit{Table \ref{tab:expertise_success_rate} shows the success rates of participants with different levels of programming expertise and Table \ref{tab:u5-a5} shows how participants' years of Python experience influenced their task completion time. Participants using \ping perform better than those using the other two tools in hard tasks in terms of both success rates and average completion time. For easy and medium tasks, \ping and GitHub Copilot show similar success rates, both significantly outperforming Multi-Turn. \ping also saves much time over the other two tools in medium tasks.} Surprisingly, we observed that programmers with 1-2 years of Python experience completed tasks faster using GitHub Copilot than with \ping. This result suggests two potential reasons. One is that our tool might not be as beneficial for novices, who may struggle with reading code comments and understanding complex logic in natural language. The other one can be the small sample size of only two participants with 1-2 years of Python experience, which could make this result a coincidence. Overall, compared to programmers with less programming experience, those with longer programming experience solved the tasks faster across the three code generation tools, which is consistent with our intuition. 

\begin{table}[t]
\centering
\caption{\edit{The success rates across three difficulty levels}}
\label{tab:difficulty_success_rate}
\begin{tabular}{c|ccc}
\toprule
{} & {{~~\textit{PInG}~~}} & {\textit{GitHub Copilot}} & {\textit{Multi-Turn}}  \\
\midrule
Easy & {4/4} & {4/4} & {2/4} \\ 
Medium & {4/4} & {3/4} & {1/4} \\ 
Hard & {2/4} & {1/4} & {0/4} \\
\bottomrule
\end{tabular}
\end{table}

\begin{table}[t]
\centering
\caption{\edit{The average task completion time for each difficulty level}}
\label{tab:difficulty_completion_time}
\begin{tabular}{c|ccc}
\toprule
{} & {{~~\textit{PInG}~~}} & {\textit{GitHub Copilot}} & {\textit{Multi-Turn}}  \\
\midrule
Easy & {10.5} & {12.2} & {17.6} \\ 
Medium & {15.0} & {17.8} & {19.4} \\ 
Hard & {18.4} & {19.1} & {20} \\
\bottomrule
\end{tabular}
\end{table}

\begin{table}[t]
\centering
\caption{The success rates of participants with different levels of programming expertise}
\label{tab:expertise_success_rate}
\begin{tabular}{c|ccc}
\toprule
{} & {{~~\textit{PInG}~~}} & {\textit{GitHub Copilot}} & {\textit{Multi-Turn}}  \\
\midrule
1-2 Years & {1/2} & {1/2} & {0/2} \\ 
2-5 Years & {4/5} & {4/5} & {1/5} \\ 
Over 5 Years & {5/5} & {3/5} & {2/5} \\
\bottomrule
\end{tabular}
\end{table}

\begin{table}[t]
\centering
\caption{The average task completion time (in minutes) of participants with different levels of programming expertise}
\label{tab:u5-a5}
\begin{tabular}{c|ccc}
\toprule
{} & {{~~\textit{PInG}~~}} & {\textit{GitHub Copilot}} & {\textit{Multi-Turn}}  \\
\midrule
1-2 Years & {17.8} & 17.5 & 20 \\ 
2-5 Years & {14.7} & 16.1 & 19.5 \\ 
Over 5 Years & {13.3} & 16.2 & 18.1 \\
\bottomrule
\end{tabular}
\end{table}

\section{Limitation and Future Work} \label{sec:discussion}

The results of our experiments demonstrate the potential for grounding-based interaction to significantly improve the accuracy and reliability of neural code generation models. Our approach, which leverages user feedback through inline comment editing, fosters a collaborative process of iterative alignment between the model's output and the user's intent. It underscores the synergy of human insights and model capabilities and thus makes code generation more reliable and adaptable through mutual grounding.

An interesting area for further analysis is studying the patterns in user edits that prove most effective for code refinement. Certain types of edits to comments, such as specifying additional conditions, correcting logical errors, or clarifying algorithm steps, may be particularly influential in guiding the model towards more accurate code refinement. Identifying and categorizing these high-leverage edits could inform techniques for eliciting more targeted feedback from users.

The comment generation and editing interface could also be enhanced to further optimize the grounding process. For example, highlighting model uncertainty and providing editing guidance could help users identify high-impact refinements more easily. Optimizing the cycle time between edits and regeneration could also affect overall productivity.

\edit{When applying \ping to complex codebases, generating inline comments for each statement may make complex code look more overwhelming. This could hinder code readability, counter to clean code principles that emphasize simplicity and minimalism in annotations~\cite{martin2009clean}. However, compared to simple code, complex code also benefits more from having detailed comments to facilitate program comprehension. We propose a couple of solutions to mitigate the negative effect of generating statement-level comments. First, we can allow users to hide all comments by default and only display the comments for the statements they do not understand. Second, we can develop a more advanced method that groups closely related statements to a block and only generates one comment for that block to reduce the number of comments.}

\edit{In addition, our user study did not differentiate between participants who were previously familiar with the assigned programming tasks and those encountering them for the first time. Given that task familiarity could influence cognitive load and problem-solving strategies, as noted by Barke et al.~\cite{barke2023grounded}, future work should incorporate this variable.}

\section{Threats to Validity} \label{sec:threats-to-validity}

A threat to internal validity is that our user feedback collected in the simulated user study was constructed by two experienced developers. The patterns in refinements and resulting accuracy improvements on this synthetic dataset may not fully reflect real-world usefulness. Our user study provides some mitigation by demonstrating productivity improvements with real users. However, the study was limited in scale and duration. More rigorous in-situ analysis is required to ascertain long-term productivity over continued tool usage.

\edit{Additionally, in our user study, we did not schedule breaks between tasks for participants. Executing tasks continuously without breaks may lead to fatigue among participants, potentially affecting their performance and the internal validity of our findings. Future studies should consider incorporating adequate breaks or spreading tasks across multiple sessions.}

In terms of external validity, we only experimented with Python code generation, which may not generalize to other languages. Our set of models was also not exhaustive, so the benefits of grounding-based interaction for other model architectures are not fully characterized. The approach may be more or less effective for very small or very large models.

\section{Related Work} \label{sec:related-work}

In recent years, large language models have demonstrated unprecedented performance on code generation tasks~\cite{chen2021evaluating, austin2021program, fried2022incoder, nijkamp2022codegen, li2023starcoder, luo2023wizardcoder, wei2023magicoder, guo2024deepseek}. While these models share similar transformer architectures, they vary in training objectives, training datasets, and model sizes. For instance, Codex~\cite{chen2021evaluating} is pre-trained on general text corpora like other GPT models and then fine-tuned on 54 million public software repositories hosted on GitHub. StarCoder~\cite{li2023starcoder} is pre-trained on a multi-lingual code corpus called Stack~\cite{Kocetkov2022TheStack} and then fine-tuned on a Python-only code corpus. Furthermore, compared with Codex, which only uses next token prediction as the training objective, StarCoder also utilizes an additional training objective called fill-in-the-middle~\cite{bavarian2022efficient}. More recently, Magicoder~\cite{wei2023magicoder} demonstrates that using LLMs to generate code instructions based on open-source code and then using them to fine-tune LLMs can significantly improve their code generation capability. 

Among the various approaches of improving LLMs for code generation, the most related to us are the prompting methods for code generation and refinement~\cite{yao2022react, yao2023tree, wang2023rap, wang2023rap, zhang2023self, jiang2023self, chen2023teaching, chen2023improving, li2023enabling, le2023codechain}.
For instance, Self-Debugging~\cite{chen2023teaching} prompts LLMs to perform iterative debugging on the generated code based on test execution results and code explanations. However, unlike our approach, they generated high-level code explanations rather than inline code comments at the statement level. Our evaluation shows that inline comments can serve as a more effective method for grounding developer intent for code refinement, even when unit tests are not available. 

In addition to test cases and code explanations, Nijkamp et al.~\cite{nijkamp2022codegen} proposed a multi-step code generation paradigm where developers can express their intent step by step in a multi-turn dialogue with an LLM. Some approaches aim to automate this by using the same LLM for task decomposition or planning~\cite{jiang2023self, yao2022react, wang2023rap, yao2023tree}. In these prompting paradigms, a developer or an agent needs to proactively decompose a task into smaller tasks to more precisely guide an LLM for code generation. By contrast, our work focuses on bi-directional communication where an agent explains the code to a developer, a developer pinpoints which part of the code is wrong, and the agent refines the erroneous part of the code accordingly.

Overall, our work differs from existing techniques by utilizing inline comments as a fine-grained grounding mechanism for code refinement. Furthermore, we have also developed specialized models for code comment generation and code refinement, rather than relying on the latent capabilities of LLMs for these specialized tasks. Our evaluation shows that developing such specialized models via fine-tuning is necessary given the distribution shift between the pre-training dataset and the targeted tasks.

\section{Conclusion} \label{sec:conclusion}

In this work, we introduced an interactive pipeline to facilitate the grounding of code generation models to user intent through inline comment editing, which helps direct the code regeneration process to better match user expectations. Experiments with multiple code generation models on two popular benchmarks reveal notable improvements in code generation accuracy when performing grounding-based code refinement. Moreover, a user study shows productivity and usability benefits when compared to alternative code generation paradigms. Our approach enhances code generation to be more aligned with user intent, making the process more manageable. It highlights the synergy between human insights and models for improving code generation, suggesting future enhancements in human-AI collaboration research.

\section{Acknowledgements} \label{sec:acknowledgements}

We thank the anonymous reviewers for their insightful feedback, as well as for the considerable time and effort they spent reviewing our work. We also thank the participants of the user studies for their valuable contributions and comments. This work was supported in part by NSF grants ITE-2333736 and CCF-2340408.

\bibliographystyle{IEEEtran}
\bibliography{custom}

\end{document}